\documentclass[cite]{epl}

\usepackage{graphicx} 
\usepackage{latexsym}
\usepackage{amssymb}
\usepackage{amsmath}

\usepackage{amssymb}
\begin{document}
\title{The design of a thermal rectifier.}

\author{Michel Peyrard\inst{1}}

\institute{\inst{1}Laboratoire de Physique, Ecole Normale Sup\'erieure
  de Lyon,\\ 
46 all\'ee d'Italie, 69364 Lyon Cedex 07, France}


\maketitle

\begin{abstract}
The idea that one can build a solid-state
device that lets heat flow more easily in one way than in the other,
forming a heat valve, is counter-intuitive. However
the design of a thermal rectifier can be easily understood
from the basic laws of heat conduction. Here we show how it
can be done. This analysis exhibits several ideas that could in
principle be implemented to design a thermal rectifier, by selecting
materials with the proper properties. In order to show the feasibility
of the concept, we complete this study by introducing a  simple model
system that meets the requirements of the design. 

\end{abstract}

While electronics has been able to 
control the flow of charges in solids for decades, 
the control of heat flow still seems out of reach, and this is why,
when a paper showed for the first time how to build a ``thermal
rectifier'' \cite{Terraneo}, the thermal analogue of the electrical
diode, it attracted a great deal of attention
\cite{RectifStories}. The idea that one can build a solid-state
device that lets heat flow more easily in one way than in the other,
forming a heat valve, is counter-intuitive and may even appear in
contradiction with thermodynamics. Actually this is not the
case, and the design of a thermal rectifier can be easily understood
from the basic laws of heat conduction. Here we show how it
can be done. This analysis exhibits several ideas that could in
principle be implemented to design a thermal rectifier, by selecting
materials with the proper properties. In order to show the feasibility
of the concept, we complete this study by introducing a  simple model
system that meets the requirements of the design. Such devices could
be useful in nanotechnology, and particularly to control he heat flow
in electronic chips.

\begin{figure}[h!]
  \begin{center}
    \includegraphics[width=6cm]{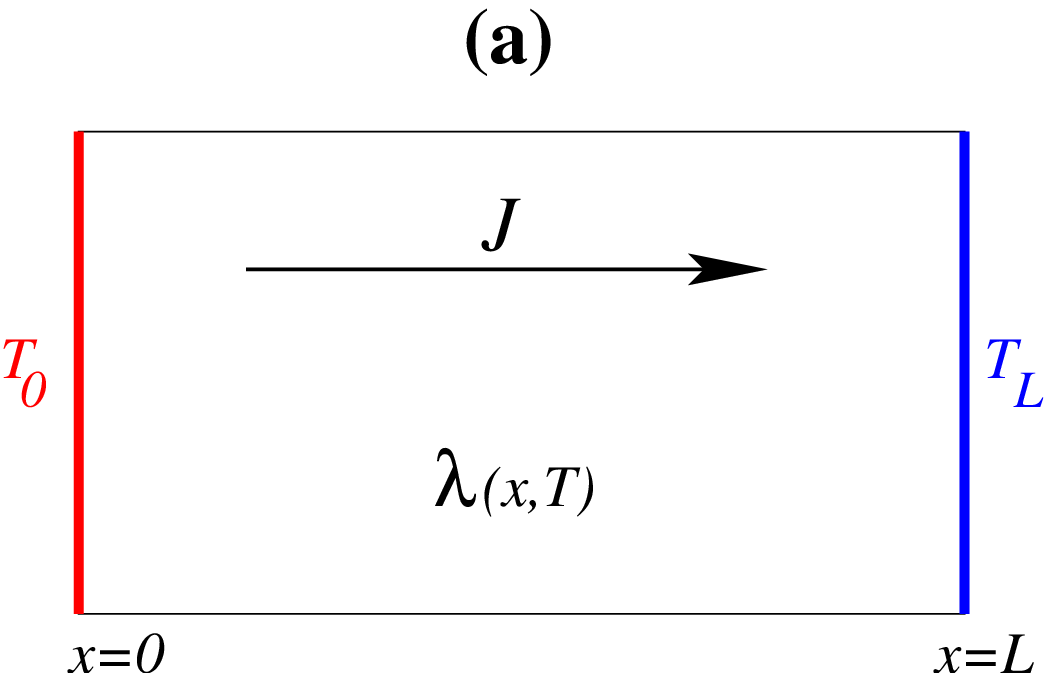} 
    \includegraphics[width=6.5cm]{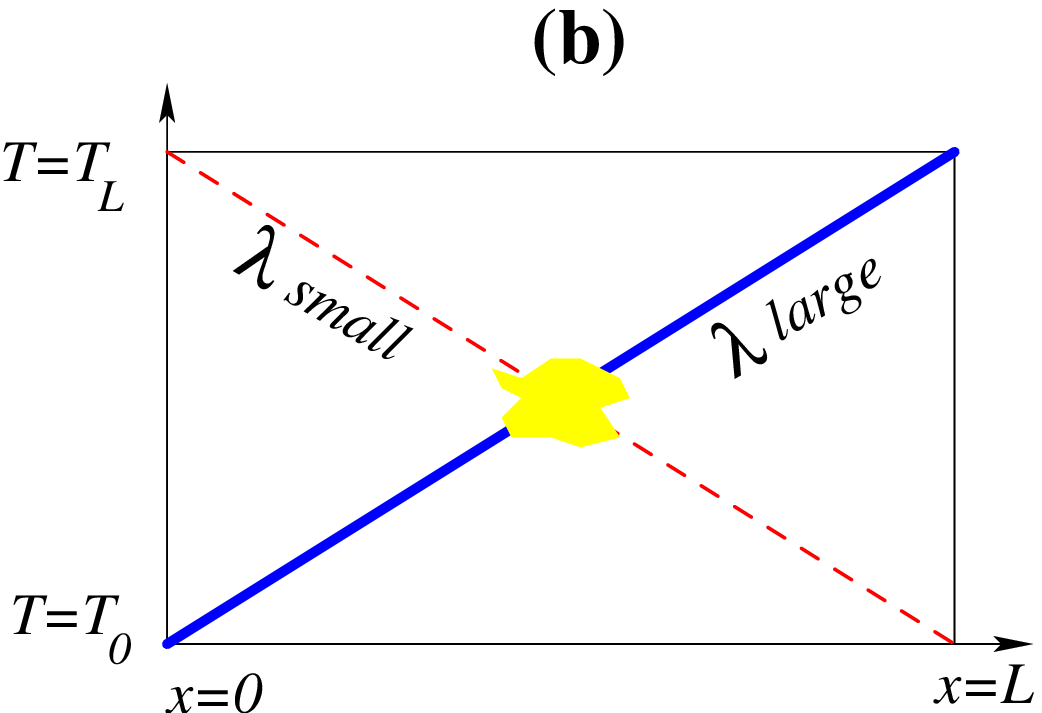}
  \end{center}
\caption{(a) Schematic picture of the device. (b) Schematic view of
  the requests on $\lambda(x,T)$ to get a rectifier.}
\label{fig:schema}
\end{figure}

\bigskip
Let us consider the heat flow along the $x$ direction, in a material
in thermal contact with two different heat baths at temperatures $T_1$
for $x=0$ and $T_2$ for $x=L$ (Fig.~\ref{fig:schema}-a). We consider
the general case of an inhomogeneous material with a local thermal
conductivity $\lambda(x,T)$ which depends not only on space but also
on temperature. To discuss the main ideas only the
$x$ dependence is introduced, but the same analysis can be extended to
a more general case, at the expense of heavier calculations.

The heat flow $J_f$ is given by
\begin{equation}
  \label{eq:flow}
J_f = - \lambda[x,T(x)] \dfrac{dT(x)}{dx} \; ,
\end{equation}
so that the temperature distribution is
\begin{equation}
  \label{eq:tdistrib}
T(x) = T_1 + \int_0^x \dfrac{J_f}{\lambda[\xi,T(\xi)]} \; d \xi \; .
\end{equation}
Solving this equation with the boundary condition $T(x=L) = T_2$
determines the value of $J_f$. A numerical solution of
Eq.~(\ref{eq:tdistrib}) can be obtained by an iterative scheme,
starting from an initial temperature distribution $T_i(x)$ which is
inserted in the r.h.s.\ of the equation to get an updated distribution
from the l.h.s., and repeating the process until the desired
convergence is achieved. A simple linear temperature distribution is a
good starting point for $T_i(x)$ and, for non-singular thermal
conductivities $\lambda(x,T)$ the convergence is fast.

If the boundary conditions are reversed, imposing temperature $T_2$
for $x=0$ and temperature $T_1$ for $x=L$, the same process leads to
another temperature distribution, and a different flux $J_r$. The
rectifying coefficient can be defined as
\begin{equation}
  \label{eq:defR}
R = \left| \dfrac{J_r}{J_f}\right| \; .
\end{equation}
In general, for arbitrary $\lambda(x,T)$, there is no condition that
imposes that $R$ should be unity.

\bigskip
From this analysis, it is easy to understand qualitatively how a
thermal rectifier can be built by selecting the appropriate function
$\lambda(x,T)$, as shown in Fig.~\ref{fig:schema}-b. In the $(x,T)$
plane, the temperature distributions for the forward and reverse
boundary conditions follow roughly the two diagonals
(Fig.~\ref{fig:schema}-b). Of course, as shown by the exact results
presented below, this is only an approximation, but it is nevertheless
sufficient to provide a guide for the choice of $\lambda(x,T)$. If the
thermal conductivity is large when the point $(x,T)$ lies on one
diagonal and small along the other, the forward and reverse heat flows
will be significantly different, i.e. the device will rectify the heat
flow. Strictly speaking the two conditions are not compatible in the
centre, but by choosing an intermediate value of $\lambda$ when
$x = L/2$ and $T = (T_1 + T_2)/2$, the rectifying effect is preserved.

\begin{figure}[h!]
  \begin{center}
    \includegraphics[width=6cm]{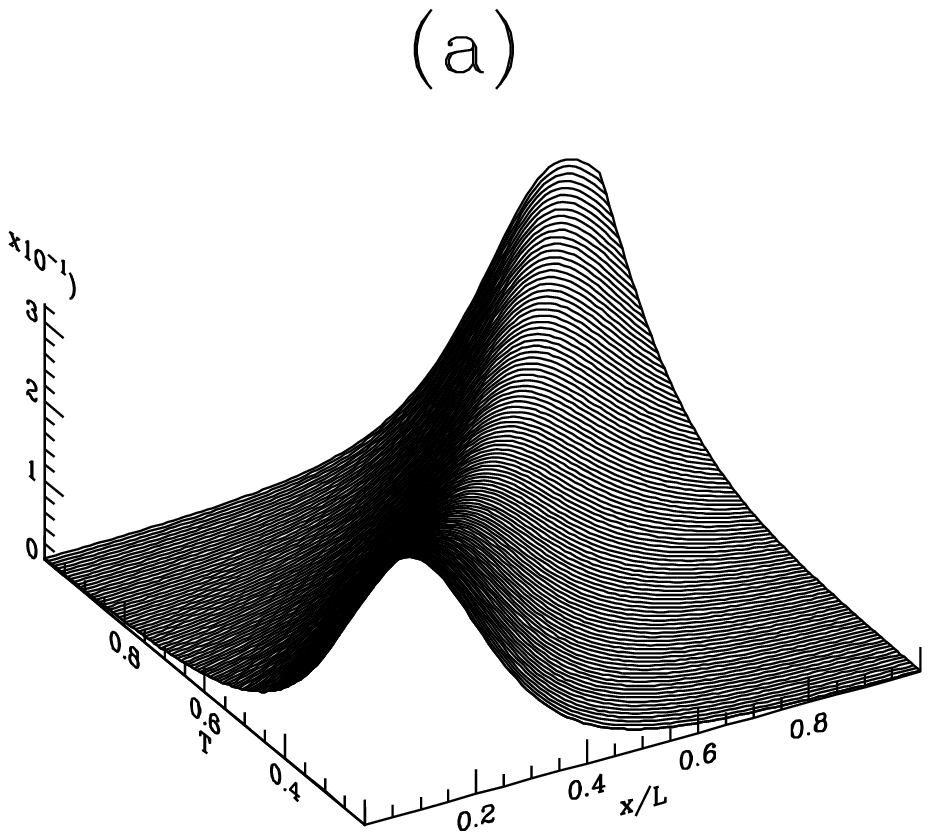} 
    \includegraphics[width=6.5cm]{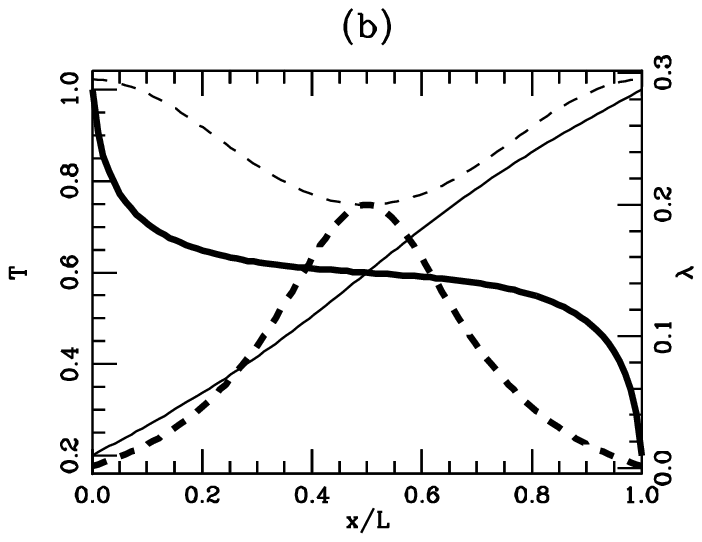} 
    \includegraphics[width=6.5cm]{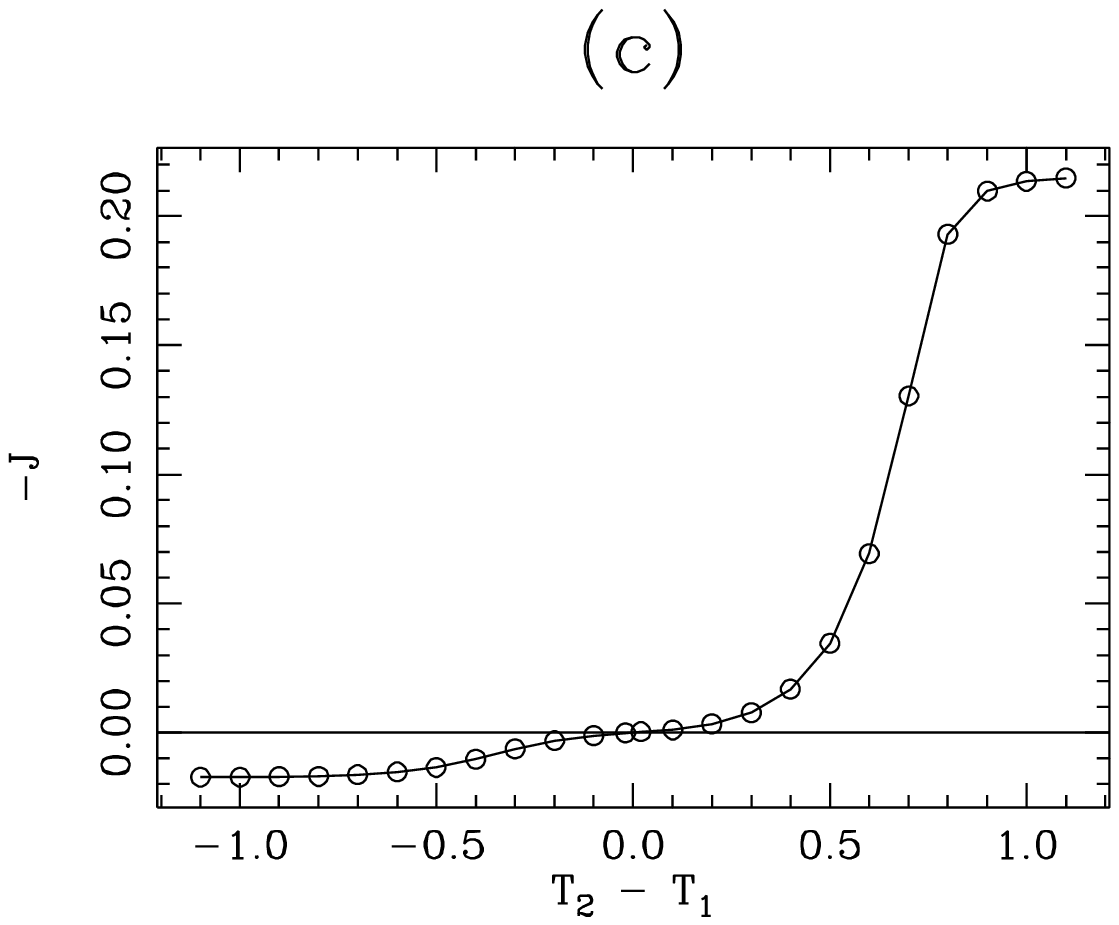}
  \end{center}
\caption{Example of temperature distributions for a particular choice
  of the thermal conductivity coefficient $\lambda(x,T) = 
\dfrac{1 - 0.8/\cosh^2(L/2-x)}
{\cosh \Big[ a (L/2 - x) + b \big(T - (T_1 + T_2)/2 \big) \Big]}$
  ($a=7$, $b=1$).
The boundary temperatures are $T_1 = 1.0$ and $T_2 = 0.2$, in arbitrary scale.
(a) Variation of $\lambda(x,T)$. (b) The temperature distributions
  (solution of Eq.~(\ref{eq:tdistrib}))
  (full lines) and the variation versus space of the local
  conductivity $\lambda[x,T(x)]$ (dashed lines) are shown 
  for the forward boundary condition
  ($T(x=0) = T_1$, $T(x=L) = T_2$) (thick lines) and reverse boundary
  condition (thin lines). For this choice of $\lambda(x,T)$, the
  rectifying coefficient is $J = |J_r/J_f|=11.5$.
(c) Variation of the heat flow $J$ across the system, as a function of
  the temperature difference $T_2 - T_1$ for a fixed value of $T_1 = 0.2$.
}
\label{fig:tdist1}
\end{figure}

Figure \ref{fig:tdist1}-a shows an example for a particular choice of
the local thermal conductivity coefficient $\lambda(x,T)$ which meets
this requirement. Solving Eq.~(\ref{eq:tdistrib}) with the forward and
reverse boundary conditions for this choice of
$\lambda$ gives temperature distributions $T_f(x)$ and $T_r(x)$, which
are not invariant by the reversal of the $x$ axis although the boundary
conditions have such a symmetry. As a result the system shows a strong
rectifying effect: the reverse heat flux is one order of magnitude
larger than the forward flux, for our choice of boundary
conditions. Figure \ref{fig:tdist1}-c shows the variation of the reverse
and forward flux as a function of the temperature difference $T_2 -
T_1$ for a given value of $T_1$. 
It has a shape that reminds the characteristic of an electrical
diode, except for high $|T_2 - T_1|$ where a saturation appears
because $\lambda(x,T)$ decreases in this range of temperatures.

The case presented on Fig.~\ref{fig:tdist1} looks of course artificial
because we have chosen a peculiar distribution $\lambda(x,T)$ to meet
the requirements that were suggested by our
analysis of Eq.~(\ref{eq:tdistrib}). Before considering a model system
that could be the basis of a solid-state rectifier, let us examine
some simpler designs, that would be more practical to build.

\begin{figure}[h!]
  \begin{center}
    \includegraphics[width=6cm]{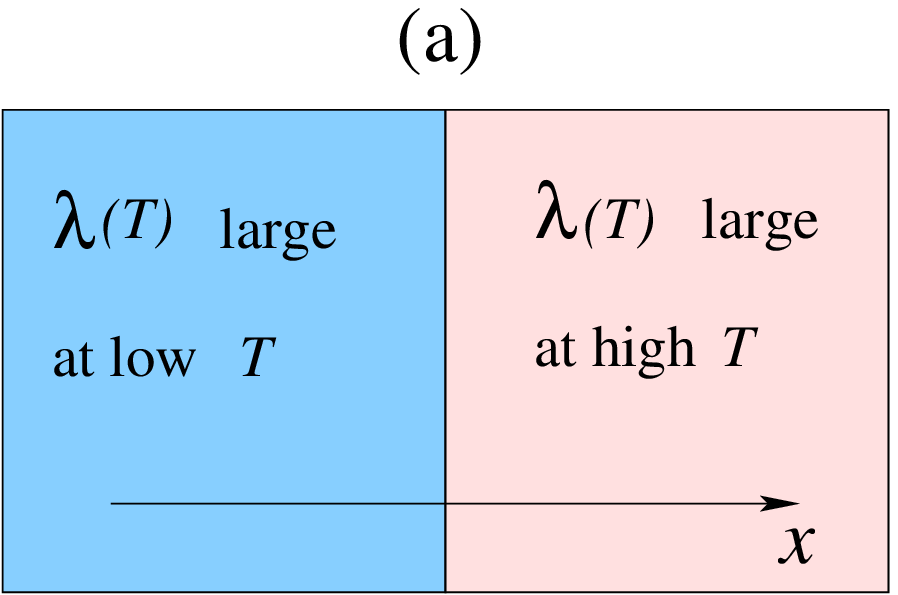} 
    \includegraphics[width=6.5cm]{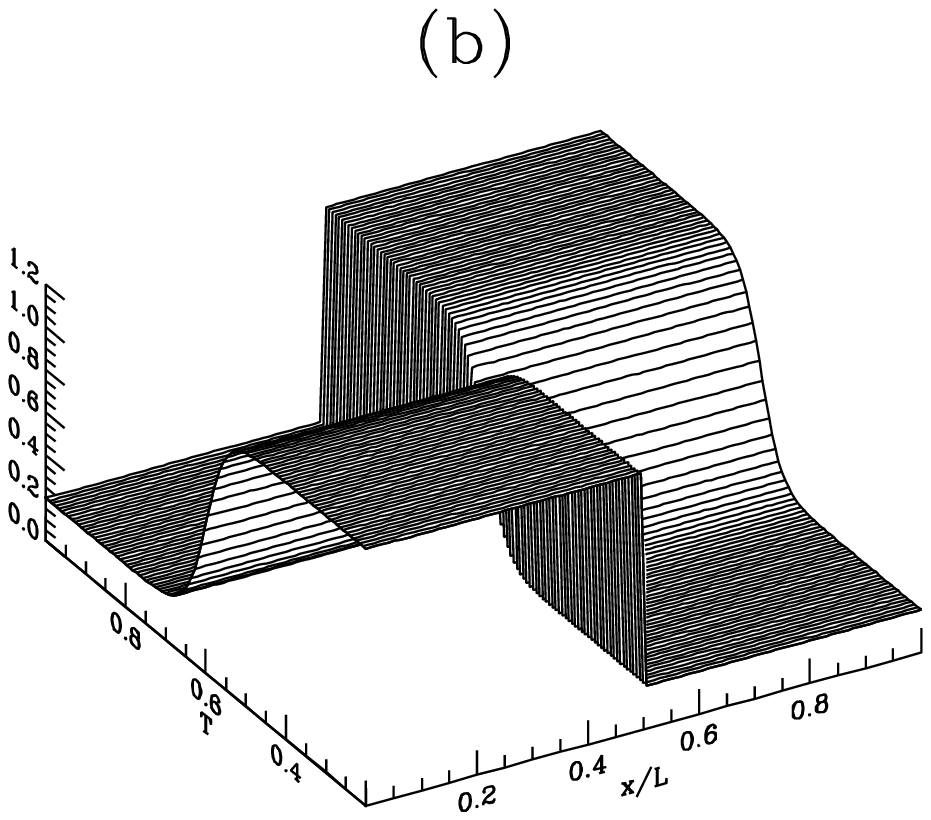} 
    \includegraphics[width=6.5cm]{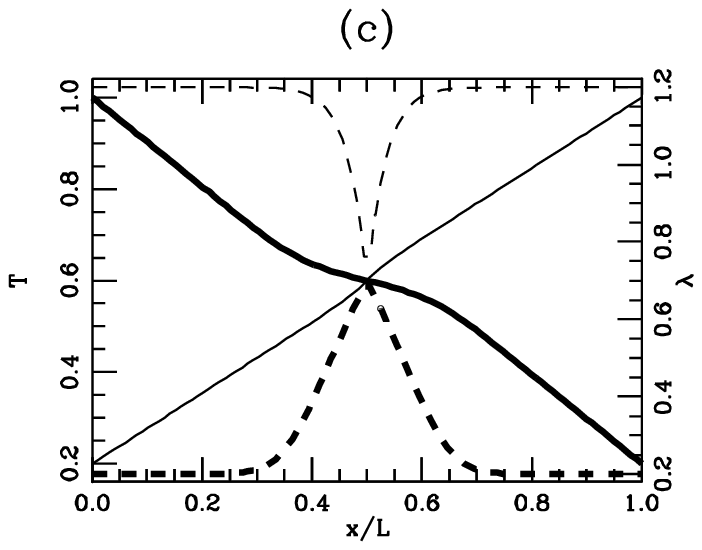}
  \end{center}
\caption{Thermal rectifier made by the juxtaposition of two different
  homogeneous materials which have a thermal conductivity that depends
  highly on temperature.
The boundary temperatures are $T_1 = 1.0$ and $T_2 = 0.2$ in arbitrary
  scale.
(a) Schematic view of the device.
(b) Variation of $\lambda(x,T)$. (c) The temperature distributions
  (solution of Eq.~(\ref{eq:tdistrib}))
  (full lines) and the variation versus space of the local
  conductivity $\lambda[x,T(x)]$ (dashed lines) are shown 
   for the forward boundary condition
  ($T(x=0) = T_1$, $T(x=L) = T_2$) (thick lines) and reverse boundary
  condition (thin lines). For this choice of $\lambda(x,T)$, the
  rectifying coefficient is $J = |J_r/J_f|=4.75$.
}
\label{fig:tdist2}
\end{figure}

One possibility is to juxtapose two homogeneous materials with thermal
conductivities $\lambda_1(T)$ and $\lambda_2(T)$ that have opposite
behaviours, $\lambda_1(T)$ decreasing sharply around some temperature
$T_c$ and $\lambda_2(T)$ rising sharply around the same temperature,
as shown in Fig~\ref{fig:tdist2}. As shown in this figure this ensures
that the average heat conductivity is higher along one diagonal of the 
$(x,T)$ plane than along the other and this can lead to a significant
value of the rectifying coefficient, which of course depends on the
exact value of $\lambda(x,T)$ and the temperatures at which the device
is operated.

\begin{figure}[h!]
  \begin{center}
    \includegraphics[width=6.5cm]{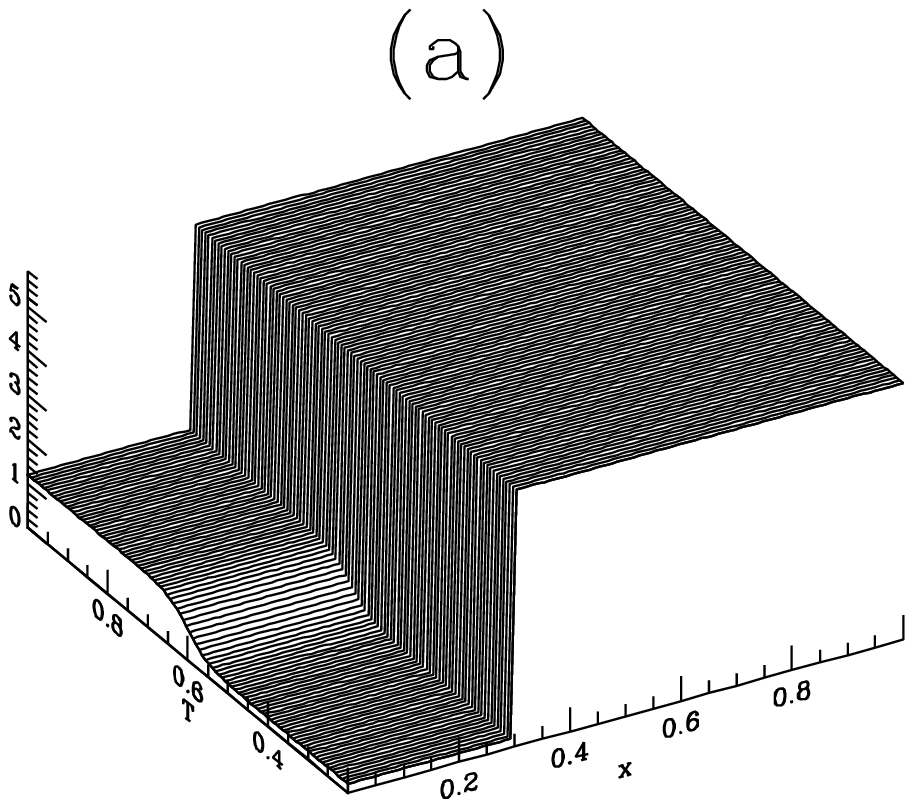} 
    \includegraphics[width=6.5cm]{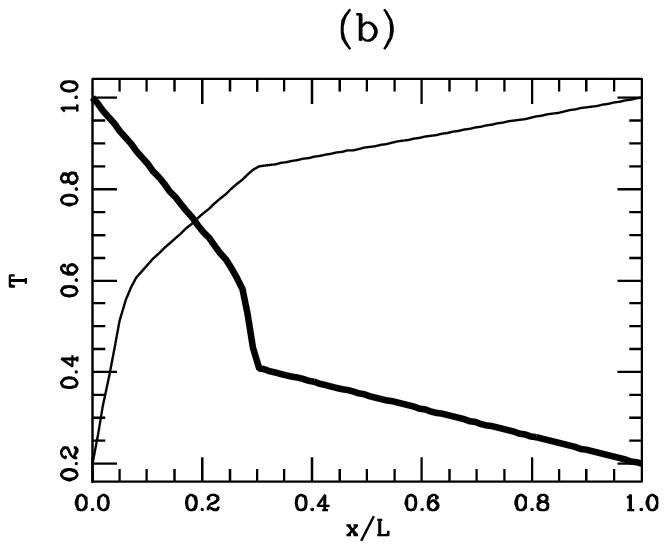}
  \end{center}
\caption{Thermal rectifier made by the juxtaposition of two different
  homogeneous materials, one of which has a thermal conductivity that
  drops below some temperature.
The boundary temperatures are $T_1 = 1.0$ and $T_2 = 0.2$ in arbitrary
  scale.
(a) Variation of $\lambda(x,T)$. (b) The temperature distributions
  (solution of Eq.~(\ref{eq:tdistrib}))
 for the forward boundary condition
  ($T(x=0) = T_1$, $T(x=L) = T_2$) (thick line) and reverse boundary
  conditions (thin line). For this choice of $\lambda(x,T)$, the
  rectifying coefficient is $J = |J_f/J_r|=1.37$.
}
\label{fig:tdist3}
\end{figure}

An even simpler design can be considered by combining a material having
a temperature dependent thermal conductivity $\lambda_1(T)$ with a
material that has a temperature independent conductivity
$\lambda_2$, as shown in Fig.~\ref{fig:tdist3}. Anticipating on the
results of the model that we discuss below, we have chosen $\lambda_2
\gg \lambda_1$, which would be the case for instance if the rectifier
were built by combining some composite material with moderate heat
conductivity $\lambda_1(T)$ with a good thermal conductor.
Although the
rectifying coefficient is smaller in this case which is not optimised, 
this simple design nevertheless shows a
significant rectifying effect for the heat flow.

\bigskip
Thus it appears that designing the simplest thermal rectifier only
requires a material with a thermal conductivity that varies
significantly as a function of temperature. In a previous study
\cite{Terraneo}, we showed that the nonlinearity of the vibrational
modes in a soft material can be used for this purpose. Let us
illustrate this with a simple one-dimensional model system. We
consider a one-dimensional lattice of nonlinear oscillators described
by the Hamiltonian
\begin{equation}
\label{eq:hamilton}
H = \sum_{n=1}^N \dfrac{p_n^2}{2m} + \dfrac{1}{2} C (y_n - y_{n-1})^2 +
V_n(y_n) \; ,
\end{equation}
where $y_n$ designates the position of particle $n$, $p_n$ its
momentum $m$ its mass, and $V_n(y_n) = D_n [\exp(- a_n y_n) - 1]^2$ is
a nonlinear on-site potential. The heat flow is carried by the phonon
modes of the lattice. In a nonlinear lattice the effective frequency
of these modes depends on their amplitude, i.e.\ on the local
temperature. An approximate calculation of the frequency shift of the
modes can be made in the self-consistent phonon approximation
\cite{Dauxois}. If one considers a lattice made of homogeneous domains
in which the parameters of the potential, $D_n$, $a_n$, are constant,
the thermal conductivity is essentially controlled by the matching of
the phonon bands at the boundaries of the domains. As the bands 
depend on temperature due to nonlinearity, the thermal conductivity
$\lambda$ depends on temperature. 

\begin{figure}
\begin{center}
  \includegraphics[height=4cm]{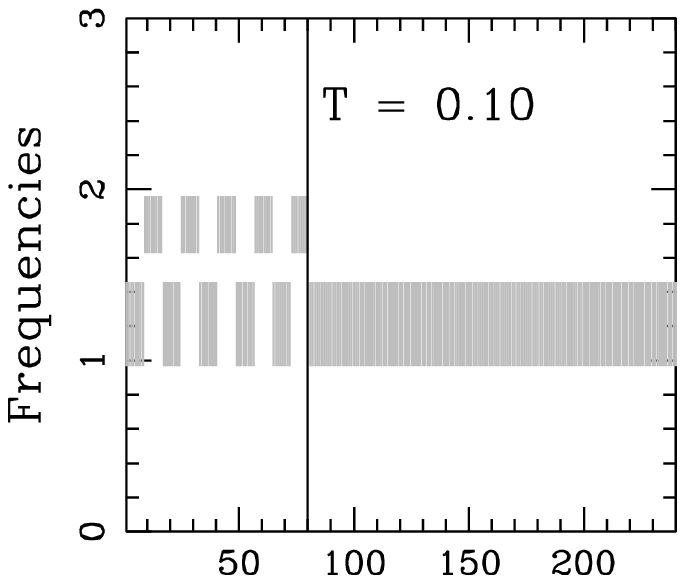} \hspace{1cm}
\includegraphics[height=4cm]{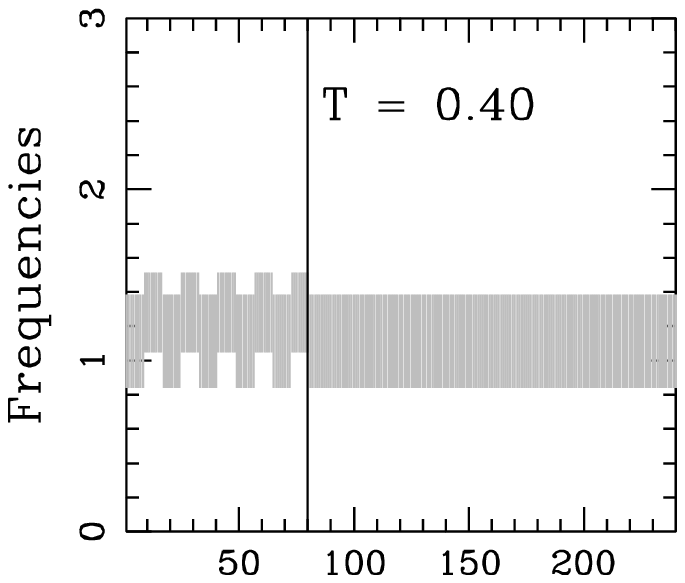}
\end{center}
\caption{Phonon bands obtained in the self-consistent phonon
  approximation for a nonlinear lattice described by Hamiltonian
  (\ref{eq:hamilton}) with 240 sites, at two temperatures (in energy
  units). The lattice is divided in two regions. In the
  right-hand-side region ($n > 80$) the lattice is homogeneous with
  $D_n = D = 0.50$ and $a_n = a = 1.0$. In the left hand side region
  ($1 \le n \le 80$) the lattice is made of striped of 8 sites each,
  in which two parameter sets for the nonlinear potential alternate,
$D_n = D = 0.5$, $a_n = a = 1.0$ on one hand, and $D_n = D' = 0.375$,
  $a_n = a' = 2.0$ on the other hand.}
\label{fig:bandes}
\end{figure}

\begin{figure}
\begin{center}
\begin{tabular}{cc}
\textbf{(a)} & \textbf{(b)} \\
\includegraphics[height=4cm]{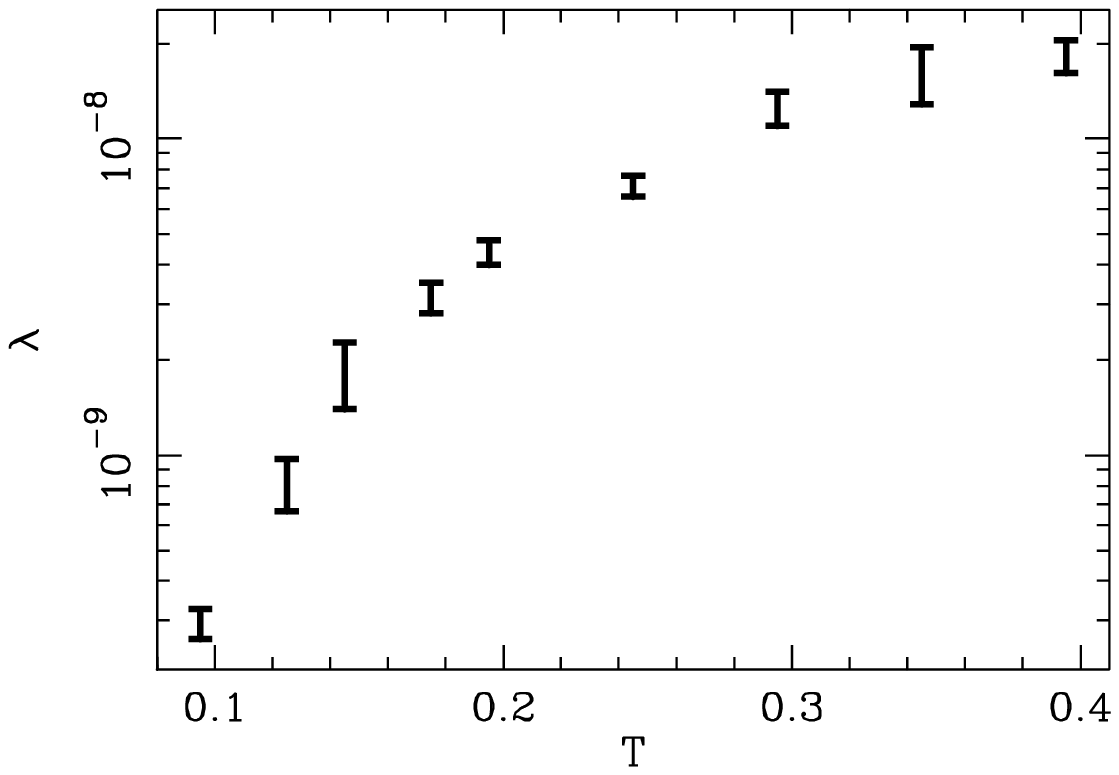}
&
\includegraphics[height=4cm]{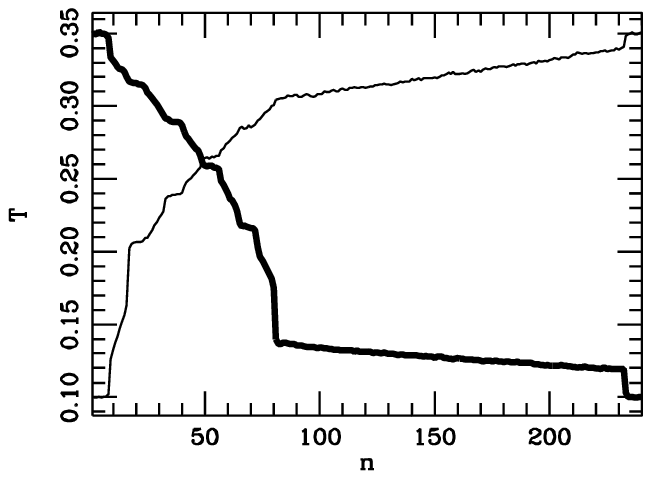} 
\end{tabular}
\end{center}
\caption{a) Variation versus temperature of the thermal conductivity of
  the composite nonlinear lattice making the left-hand-side region of
  the model introduced in Fig.~\ref{fig:bandes}.
b) Temperature distributions observed in the simulation of the
  nonlinear lattice described in the caption of Fig.~\ref{fig:bandes}
  in contact with two thermal baths at temperatures $T_1 = 0.1$ 
and $T_2 = 0.35$. The rectifying coefficient deduced from the
  simulation is $R = |J_f/J_r| = 1.39$.
}
\label{fig:lambdaT}
\end{figure}

Figure \ref{fig:bandes} shows the self-consistent phonon bands in
a nonlinear lattice made of two
different regions. In the homogeneous right-hand-side region, the
frequencies of the vibrational modes are only 
weakly temperature dependent and do not vary in space. Phonons
propagate easily in this region which has a high heat conductivity
$\lambda_2$ which is almost temperature independent. In the
left-hand-side region, the phonon bands of the two types of stripes do
not match at low temperature. The discontinuities scatter phonons and
the heat conductivity $\lambda_1$ is low at low 
temperature ($T = 0.1$). But, as $T$ is raised, the self-consistent
phonon frequencies in the stripes with the highest nonlinearity ($a' =
2.0$) drop and overlap with the frequencies in the other stripes. The
thermal conductivity $\lambda_1$ raises sharply at high temperature
($T=0.4$). This analysis based on self consistent phonon calculations
is confirmed by numerical simulations in which the ends of a lattice
similar to the left-hand-side region are thermalized by Langevin
thermostats at two slightly different temperatures $T$ and $T + \Delta
T$ with $\Delta T = 0.01$. The value of $\lambda_1(T)$ is then estimated
by measuring the heat flux $J(T)$ across the lattice \cite{Terraneo} and
using $\lambda_1(T) = J(T) \; \Delta x / \Delta T$ where $\Delta x$ is the
length of the lattice segment used for the measurement (132 sites).
Figure~\ref{fig:lambdaT}-a shows the result of this numerical
measurement of $\lambda_1(T)$ and Fig.~\ref{fig:lambdaT}-b shows
the forward and reverse temperature distributions in the nonlinear
lattice in contact with two thermal baths at temperatures $T_1 = 0.1$
and $T_2 = 0.35$. These curves are very similar to the theoretical
curves of
Fig.~\ref{fig:tdist3}-b, which indicates 
that the nonlinear lattice is able to
lead to the kind of thermal conductivity distribution $\lambda(x,T)$
shown in Fig.~\ref{fig:tdist3}-a. It behaves as a thermal rectifier.

\bigskip
In conclusion, we have shown that a simple calculation of the heat
flowing through a material which has a thermal conductivity that
depends on space {\em and} temperature indicates that thermal
rectifiers, in which the heat flows more easily in one direction than
in the other, can be designed. A nonlinear lattice models,
investigated by numerical simulation, confirms the validity of the
analysis and shows that materials meeting the necessary requirements
to build thermal rectifiers could exist. This nonlinear lattice can be
viewed as a simplified description of a layer of soft molecules
deposited on a substrate which has a very small thermal conductivity
such as a glass. But, since our analysis shows that any material which
has a temperature dependent thermal conductivity can be used, one can
think of other possibilities such as composite materials or solids in
the vicinity of a phase transition. 

Moreover, in a practical design one could take advantage of another
degree of freedom that we have not exploited in our simple analysis:
geometry. For instance changing the width of the conducting layer is
an easy way to control the spatial dependence of $\lambda(x,T)$. We
have verified that this allows the design of a thermal rectifier made
of a single homogeneous material, having a temperature dependent
thermal conductivity $\lambda(T)$ by selecting the proper shape of the
conducting layer.

The simplicity of the concept suggests that, after the selection of
the optimal material, it could be used in various applications in
nanotechnology, such as the control of the heat flow in electronic
chips for instance.

\end{document}